\documentclass[preprint,authoryear,12pt]{elsarticle}
\usepackage[a4paper]{geometry}

\usepackage{amsmath}
\usepackage{amssymb}
\usepackage{graphicx}
\usepackage{lineno}
\usepackage{natbib}
\usepackage{color}

\newcommand{\dd}[2]{\frac{\mathrm{d} #1}{\mathrm{d} #2}}
\newcommand{\dpar}[2]{\frac{\partial #1}{\partial #2}}
\renewcommand{\vec}[1]{\mathbf{#1}}
\newcommand{\svec}[1]{\boldsymbol{#1}}
\newcommand{\tensor}[1]{\mathsf{#1}}

\graphicspath{ {Images/} }

\begin{document}

\begin{frontmatter}

\title{Bayesian inference of ocean diffusivity from Lagrangian trajectory data}

\author[ed]{Y. K. Ying\corref{yky}}

\author[ed]{J. R. Maddison}

\author[ed]{J. Vanneste}

\address[ed]{School of Mathematics and Maxwell Institute for Mathematical Sciences, The University of Edinburgh, Edinburgh, EH9 3FD, United Kingdom}
\cortext[yky]{Corresponding author (Y.K.Ying@ed.ac.uk)}

\begin{abstract}

A Bayesian approach is developed for the inference of an eddy-diffusivity field from Lagrangian trajectory data. The motion of Lagrangian particles is modelled by a stochastic differential equation associated with the advection--diffusion equation. An inference scheme is constructed for the unknown parameters that appear in this equation, namely the mean velocity, velocity gradient, and diffusivity tensor. The scheme provides a posterior probability distribution for these parameters, which is sampled using the Metropolis--Hastings algorithm.
The approach is applied first to a simple periodic flow, for which the results are compared with the prediction from homogenisation theory, and then to trajectories in a three-layer quasigeostrophic double-gyre simulation. The statistics of the inferred diffusivity tensor are examined for varying sampling interval and compared with a standard diagnostic of ocean diffusivity. The Bayesian approach proves capable of estimating spatially-variable anisotropic diffusivity fields from a relatively modest amount of data while providing a measure of the uncertainty of the estimates.
\end{abstract}


\end{frontmatter}

\section{Introduction}

Turbulent processes can lead, on sufficiently long time scales, to diffusive mixing of tracer quantities \citep{taylor1922,majda1999}. In the ocean large-scale instabilities gives rise to geostrophic eddies. These energetic eddies dominate the redistribution of heat and tracers both laterally and vertically \citep[e.g.][]{jayne2002} and contribute to the formation of large-scale circulation patterns \citep[e.g.][]{marshall2003, hallberg2006}. The mixing induced by these eddies is typically modelled through an ``eddy diffusivity''. Diffusive models can be shown to be valid in limiting cases \citep[e.g.][]{davis1987,majda1999}, although the empirically long ($\sim 100$~days) time for the diffusive regime to come into effect in some parts of the ocean \citep{rypina2012} makes their general applicability questionable.

There are multiple approaches for the diagnosis of turbulent ocean eddy diffusivities, which are not obviously equivalent. One can diagnose a diffusivity from turbulent eddy fluxes \citep[e.g.][]{bachman2013}, although this may be prone to ambiguity due to the possible presence of rotational fluxes \citep{marshall1981}. Alternatively, observations of the motion of tracer contours can be used to define an eddy diffusivity \citep{nakamura1996,marshall2006}. A separate broad class of diffusivity diagnostics is based upon observations of the motion of fluid parcels \citep[e.g.][]{lacasce2008,vansebille2018}, which may for example be obtained from simulated Lagrangian trajectories, or from ocean drifter data. For comparisons between these approaches see \citet{klocker2012b} and \citet{abernathey2013}.

Consider Lagrangian particles, where the $i$th particle has position $\vec{X}_i \left( t \right)$ and corresponding displacement $\vec{S}_i \left( t \right) = \vec{X}_i \left( t \right) - \vec{X}_i \left( 0 \right)$. In a statistically stationary and homogeneous flow one may define an absolute diffusivity based upon the absolute dispersion of particles \citep[][]{taylor1922,lacasce2008}
\begin{equation} \label{eqn:kappa_abs}
  \kappa_\mathrm{abs} \left( \tau \right) =
    \frac{1}{2} \frac{\mathrm{d}}{\mathrm{d}\tau}
    \left<
      \vec{S}_i \left( \tau \right)
      \otimes \vec{S}_i \left( \tau \right) 
    \right>,
\end{equation}
where $\left< \cdot \right>$ denotes an appropriate average over particles, such as an ensemble average, and $\tau$ represents a time window over which the particle trajectories are considered. As $\tau \to \infty$, $\kappa_\mathrm{abs}\left( \tau \right)$ converges to a constant and characterises the asymptotic growth rate of particle dispersion. This definition makes no correction for the possible presence of a background mean flow, which can for example be accounted for via
\begin{equation}
  \kappa_\mathrm{abs} \left( \tau \right) =
    \frac{1}{2} \frac{\mathrm{d}}{\mathrm{d}\tau}
    \left<
      \left( \vec{S}_i \left( \tau \right) - \left< \vec{S}_i \left( \tau \right) \right> \right)
      \otimes \left( \vec{S}_i \left( \tau \right) - \left< \vec{S}_i \left( \tau \right) \right> \right)
    \right>,
\end{equation}
correcting for a mean drift \citep[e.g.][]{sallee2008}.

Retaining the assumption of a statistically stationary and homogeneous flow, one may define a relative diffusivity \citep[e.g.][]{lacasce2008}

\begin{equation}
  \kappa_\mathrm{rel} \left( \tau \right) =
    \frac{1}{4} \frac{\mathrm{d}}{\mathrm{d}\tau}
    \left<
      \left( \vec{X}_{i_1} \left( \tau \right) - \vec{X}_{i_2} \left( \tau \right) \right)
      \otimes \left( \vec{X}_{i_1} \left( \tau \right) - \vec{X}_{i_2}\left( \tau \right) \right)
    \right>,
\end{equation}
where now the average is taken over all distinct pairs of particles ($i_1 \neq i_2$). This  automatically takes account of the presence of a uniform background mean flow. Such a relative diffusivity has been used to study energy spectra in fluid turbulence \citep[e.g.][]{koszalka2009,lumpkin2010}.

The above definitions make use of statistical homogeneity to yield a single bulk uniform diffusivity. This is problematic if the diffusivity is expected to vary in different regions of the ocean. To account for this, \citet{davis1987, davis1991} defines  the spatially dependent diffusivity 
\begin{equation}
    \kappa_\mathrm{Davis} \left( \vec{x}; \tau \right)
        = \int_{-\tau}^0
        \left<
            \left[ \dot{\vec{X}_i}\left( t \right) - \overline{\vec{u}} \left( \vec{x} \right) \right] \otimes \left[ \dot{\vec{X}_i}\left( t+s \right) - \overline{\vec{u}} \left( \vec{X}_i \left( t+s \right) \right) \right]
        \right>_{ \{ \vec{X}_i\left( t \right) = \vec{x} \} } \mathrm{d}s,
        \label{eqn:kappa_davis}
\end{equation}
where the conditional average $\left< \cdot \right>_{ \{ \vec{X}_i\left( t \right) = \vec{x} \} }$ is taken over all trajectories $\vec{X}_i\left( t \right)$ that pass through position $\vec{x}$ at some time $t$.
While this definition captures spatial variations in diffusivity, it requires the choice of an appropriate background mean flow $\overline{\vec{u}} \left( \vec{x} \right)$. Its implementation is further complicated by the need for past history information of particles which arrive at a common point -- in practice this necessitates local binning of particles which arrive in the vicinity of a point, and may also be replaced with future information of particles which leave the vicinity \citep[e.g.][]{oh2000, griesel2010, klocker2012b,ruehs2018}.

A concern in the \citet{davis1987} diffusivity is its dependence on the time-lag parameter $\tau$. One may hope for convergence in the large-$\tau$ limit,  after some characteristic decorrelation time, but this decorrelation time may be sufficiently large that the particles have left the neighbourhood of $\vec{x}$. As a result, particles involved in the calculation experience different flow regions, with different diffusivity properties, over the timescale $\tau$ over which the integral is taken. These non-local effects mean that care needs to be exercised when interpreting the spatial dependence of the \citet{davis1987} diffusivity. Further, there is the concern that in general this diffusivity need not be non-negative definite, nor even symmetric.

In this article we present a new approach for the diagnosis of ocean eddy diffusivity from Lagrangian particle data using Bayesian inference. Given a stochastic model for the particle motion, discretely observed Lagrangian particle positions, and prior information, the approach infers a joint posterior probability distribution for both a local flow velocity and a local anisotropic diffusivity tensor. 
This probability distribution makes it possible, for example, to compute mean quantities or to find maximum a posteriori estimates, and to quantify the uncertainty of these estimates.

The paper is organised as follow. In section \ref{sect:bayesian} the Bayesian inference approach and its implementation using Monte Carlo Markov Chain are described. Section \ref{sect:idealised} provides an application in an idealised configuration. In section \ref{sect:qg} the approach is applied to Lagrangian particle data obtained from a three-layer quasigeostrophic double-gyre calculation, and the resulting diffusivity diagnosis is compared against the \citet{davis1987} diffusivity. The paper concludes in section \ref{sect:conclusions} with an outlook towards more general applications of Bayesian inference to the analysis of Lagrangian drifter data.

\section{Mathematical background}\label{sect:bayesian}

\subsection{Stochastic Lagrangian particle dynamics}

The position $\vec{X} \left( t \right)$ of  particles advected in a time-dependent velocity field $\vec{u}(\vec{x},t)$ satisfies the ordinary differential equation
\begin{equation} \label{eqn:ode}
  \dd{\vec{X}}{t} = \vec{u} \left( \vec{X} \left( t \right), t \right),
\end{equation}
subject to some initial condition $\vec{X} \left( 0 \right) = \vec{x}_0$. The concept of eddy diffusivity arises when attempting to coarse-grain this equation: it might be expected that over sufficiently long time scales the behaviour of its solutions is well captured by the Markov-0 model \citep[][]{berloff2002b}, which is a stochastic differential equation (SDE)
\begin{equation}\label{eqn:sde}
  \mathrm{d}\vec{X} = \left[ \vec{U} \left( \vec{X} \left( t \right) \right) + \nabla \cdot \tensor{K} \left( \vec{X} \left( t \right) \right) \right] \mathrm{d}t
    + \sqrt{2 \, \tensor{K} \left( \vec{X} \left( t \right) \right)} \mathrm{d}\vec{W}.
\end{equation}
Now $\vec{U}$ is a time-independent average velocity field, $\tensor{K}$ is the eddy diffusivity which is a symmetric positive definite tensor (whose square root is uniquely defined by requiring that it too be symmetric positive definite), and $\vec{W}$ is multi-dimensional Brownian motion. The reduction from \eqref{eqn:ode} to the Markov-0 model \eqref{eqn:sde} can only be justified rigorously, and explicit expressions for $\vec{U}$ and $\tensor{K}$ can only be obtained, when $\vec{u}(\vec{x}, t)$ satisfies strong assumptions of scale separation in time and/or space that are not met in the context of the ocean \citep[see][and the reference therein]{griffa1996}. Here we adopt a heuristic approach and seek to estimate values for $\vec{U}$ and $\tensor{K}$ that are most consistent -- in a sense to be explained -- with a set of observed particle trajectories $\vec{X}_i(t)$. 

The evolution of $\vec{X}(t)$ according to \eqref{eqn:sde} is entirely characterised by the transition probability density $\pi \left( \vec{x}, t | \vec{x}_0 \right)$ which defines the probability of finding the particle in the neighbourhood of $\vec{x}$ at time $t$ given it is initially at $\vec{x}_0$. The transition probability evolves under the Fokker--Planck equation
\citep[e.g.][]{evans2013, pavliotis2014}
\begin{equation}\label{eqn:fp}
  \dpar{\pi}{t} + \nabla \cdot \left( \vec{U} \pi \right) = \nabla \cdot \left( \tensor{K} \nabla \pi \right),
\end{equation}
with initial condition $\pi \left( \vec{x}, 0 | \vec{x}_0 \right) = \delta \left( \vec{x} - \vec{x}_0 \right)$. This is   the advection--diffusion equation, and hence \eqref{eqn:sde} is a natural stochastic
model for advective and diffusive processes.

The velocity and diffusivity fields $\vec{U}$ and $\tensor{K}$ are fields defined over the entire spatial domain. For practical computations it is necessary to first discretise these fields over space,
\begin{equation} \label{eqn:finitedim}
  \vec{U}(\vec{x}) = \vec{U}(\vec{x} ; \svec{\theta}) \quad \textrm{and} \quad {\tensor{K}}(\vec{x}) = {\tensor{K}}(\vec{x} ; \svec{\theta}),
\end{equation}
where $\svec{\theta}$ denotes the degrees of freedom for both $\vec{U}$ and $\tensor{K}$ -- that is, $\svec{\theta}$ is a finite-length vector of parameters which specifies the discrete approximation for $\vec{U}$ and $\tensor{K}$. Hereafter the dependence of quantities on $\svec{\theta}$ is omitted, but it should be borne in mind that most objects of interest, the transition probability $\pi$ for instance, have such a dependence. The problem of estimating the discretised velocity and diffusivity fields now reduces to the estimation of $\svec{\theta}$. In the Bayesian-inference approach we adopt, $\svec{\theta}$ is regarded as a random variable and its entire probability distribution, and hence a probability distribution for $(\vec{U},\tensor{K})$, is estimated from trajectory data. 

\subsection{Bayesian inference}

Given $N$ particles each observed at $P$ distinct times $t_j$, evolving under the SDE \eqref{eqn:sde}, Bayes' theorem gives \citep[a thorough textbook reference for Bayesian statistics is][]{gelman2013}
\begin{equation}\label{eqn:bayes}
  p \left( \svec{\theta} | R \right)
    = \frac{p \left( R | \svec{\theta} \right) p \left( \svec{\theta} \right)}
      {\int p \left( R | \svec{\theta} \right) p \left( \svec{\theta} \right) \mathrm{d}\svec{\theta}}
    \propto p \left( R | \svec{\theta} \right) p \left( \svec{\theta} \right),
\end{equation}
where the integral is over the full parameter space. $R$ denotes the data, and can be set equal to the full trajectory,
\begin{equation}
  R = \left\{ \left( i, \vec{X}_i \left( t_j \right), t_j \right) :
    i =1, \cdots, N, \ j = 1, \cdots, P  \right\},
\end{equation}
where $\vec{X}_i \left( t_j \right)$ is the position of the $i$th particle at the $j$th observation time. Equivalently, as the SDE is Markovian, $R$ can be replaced with
\begin{equation}
  R = \left\{ \left( \vec{X}_i \left( t_j \right), \vec{X}_i \left( t_{j + 1} \right), t_{j + 1} - t_j \right) :
    i = 1, \cdots, N, \ j = 1, \cdots, P - 1  \right\}.
\end{equation}
That is, the data consist of the start and end positions of each particle between consecutive pairs of observations, and the time separation between the observations. Note that this is easily generalised for the case of differing observation times for each particle and differing particle trajectory lengths.

Three key probability distributions appear in \eqref{eqn:bayes}: the posterior $p \left( \svec{\theta} | R \right)$, the likelihood $p \left( R | \svec{\theta} \right)$, and the prior $p \left( \svec{\theta} \right)$. The posterior $p \left( \svec{\theta} | R \right)$ is the probability distribution of the parameter $\svec{\theta}$ given the observations and the model, and its determination is the goal of the inference. It should be interpreted as an objective measure of the plausibility of a certain value of $\svec{\theta}$ (and hence of $\vec{U}$ and $\tensor{K}$) in view of the observations, assuming the model is perfect.
The likelihood $p \left( R | \svec{\theta} \right)$ is the probability that particles evolving according to \eqref{eqn:sde}, and with $(\vec{U},\tensor{K})$ fixed by $\svec{\theta}$, have positions matching $R$. It is given explicitly in terms of a product of transition probabilities
\begin{equation} \label{eqn:likelihood}
  p \left( R | \svec{\theta} \right) = \prod_{i=1}^N \prod_{j=1}^{P-1} \, \pi \left( \vec{X}_i \left(
      t_{j + 1} \right), t_{j + 1} - t_j | \vec{X}_i \left( t_j \right) \right).
\end{equation}
The prior $p \left( \svec{\theta} \right)$ is a subjective choice for the plausibility of a given set of parameters $\svec{\theta}$ in the absence of data. Its importance for the posterior diminishes as the number $N(P-1)$ of data points increases. 

\subsection{Sampling: Metropolis--Hastings} \label{sec:samp_mh}

Assuming we can evaluate the transition probability in \eqref{eqn:likelihood}, Bayes' formula \eqref{eqn:bayes} gives the probability density for the parameters $\svec{\theta}$ and therefore for $\vec{U}$ and $\tensor{K}$ in an explicit form. This is however a probability density in a high-dimensional space which cannot be visualised and from which derived quantities cannot be computed directly. Instead, one is interested in computing integrals of various quantities against the posterior -- that is, in evaluating
\begin{equation} \label{eqn:int}
  \int f \left( \svec{\theta} \right) p \left( \svec{\theta} | R \right) \mathrm{d}\svec{\theta}
\end{equation}
for some $f \left( \svec{\theta} \right)$. For example $f \left( \svec{\theta} \right) = \tensor{K}$ yields the posterior mean diffusivity, $\bar{\tensor{K}}$ say, which can used as an estimate for the eddy diffusivity, while 
$f \left( \svec{\theta} \right) = \|\tensor{K} - \bar{\tensor{K}}\|^2$
yields a variance characterising the uncertainty of the estimate $\bar{\tensor{K}}$.

Markov Chain Monte Carlo (MCMC) methods can be used to obtain numerical approximations for integrals of the form \eqref{eqn:int}.
These methods generate sequences of random samples $\svec{\theta}^{(k)}$ using a transition probability $T(\svec{\theta}^{(k+1)}  | \svec{\theta}^{(k)})$ chosen to ensure that, for large $k$, the $\svec{\theta}^{(k)}$ are distributed 
according to $p \left( \svec{\theta} | R \right)$. The integrals \eqref{eqn:int} are then estimated simply by the arithmetic mean of $f ( \svec{\theta}^{(k)} )$. Here we use the well-known Metropolis--Hastings algorithm, based on an acceptance/rejection definition of $T(\svec{\theta}^{(k+1)} | \svec{\theta}^{(k)})$, and more specifically the Gibbs sampler \citep[e.g.][]{geman1984} for which the successive samples $\svec{\theta}^{(k)}$ and $\svec{\theta}^{(k+1)}$ differ in at most one component. 
The reliable estimation of integrals using MCMC requires monitoring the convergence of the estimates and ensuring that the $\svec{\theta}^{(k)}$ properly explore the support of $p \left( \svec{\theta} | R \right)$; we adopt the \citet{gelman1992} diagnostic (also in \ref{app:metropolis_hastings} and in section 11.4 of \cite{gelman2013}) to verify this.


\subsection{Local inference}\label{sect:local_inference}

The specific inference problem considered in this article is conducted in a local cell-wise manner. The domain of interest is partitioned into a coarse mesh, and we seek to obtain information on the flow velocity and diffusivity for each mesh cell. The result of the inference is expected to be dependent on the choice of mesh, and in particular on the mesh cell size. This is consistent with the coarse graining involved in approximating \eqref{eqn:ode} by \eqref{eqn:sde} -- the eddy diffusivity obtained is dependent upon the spatial scales.

Note that a meaningful eddy diffusivity is only realised after a decorrelation time scale. Over short time scales, correlated advection associated with the so-called ``ballistic'' regime \citep[e.g.][]{pasquero2007, rypina2012} dominates and is incompatible with the diffusive model \eqref{eqn:sde}. It is therefore necessary to ensure that the pairs of observed particle positions employed are separated by a sufficient time interval -- a principle noted in a multi-scale system in \citet{pavliotis2007} \citep[see also][for an application to eddy diffusivity]{cotter2009}. An optimal sampling interval, which discards the minimum number of position records while preserving the validity of the model \eqref{eqn:sde}, is rarely known a priori. In practice the inference is performed with varying sampling intervals and the convergence of the various estimates is examined. In the local inference approach we take here it is also necessary for the particles to remain in (or at least close to) the cell considered over the sampling interval. There is therefore a trade-off  between two competing requirements: the sampling interval must be long enough that the particles do decorrelate, and short enough that they are not transported far from the considered cell. One must therefore take care to choose an appropriate sampling interval between observations, and be aware that this may not always exist. The possibility for a more advanced ``non-local'' inference, which alleviates this difficulty, is discussed in the conclusions.

\section{Idealised example: Taylor--Green vortices with a background flow}\label{sect:idealised}

\subsection{Configuration}

A highly idealised model of oceanic eddies in a background current is constructed by superimposing a constant mean flow on top of Taylor--Green vortices, leading to the two-dimensional and doubly-periodic steady velocity field
\begin{equation} \label{eqn:TG}
  \vec{u}(\vec{x}) = U_\mathrm{TG} \left(
    \begin{array}{c} - \sin \left( {2 \pi x }/{L} \right) \cos \left( {2 \pi y }/{L} \right) \\
                       \cos \left( {2 \pi x }/{L} \right) \sin \left( {2 \pi y }/{L} \right) \end{array} \right)
                                           + U_\mathrm{M} \left(
    \begin{array}{c} \cos \phi_\mathrm{M} \\
                     \sin \phi_\mathrm{M} \end{array} \right),
\end{equation}
where $U_\mathrm{TG}$ is the maximum vortex speed, $U_\mathrm{M}$ is a background flow speed, and $\phi_\mathrm{M}$ is the angle of the background flow to the $x$-axis. The small-scale advection--diffusion of particles according to
\begin{equation} \label{eqn:ad-diff-TG}
\mathrm{d} \vec{X} = \vec{u}(\vec{X}(t)) \, \mathrm{d} t + \sqrt{2 \kappa} \, \mathrm{d} \vec{W},
\end{equation}
is considered, where $\kappa \not= 0$ is here a small-scale scalar diffusivity. Note that $\kappa$, which governs the small-scale motion of the particles, is not the object to be inferred in this problem. Rather we seek to infer information about a large-scale effective diffusivity, which governs the long-time behaviour.

Homogenisation theory \citep[e.g.][]{majda1993, majda1999} provides rigorous coarse-graining results for this problem. Specifically, over scales much larger than the vortex period $L$, the motion of particles is approximated by the SDE \eqref{eqn:sde} with a uniform mean velocity $\vec{U}=U_\mathrm{M} (\cos \phi_\mathrm{M},\sin \phi_\mathrm{M})$ and an effective diffusivity tensor $\tensor{K}$. The effective diffusivity tensor $\tensor{K}$ can be computed by solving a two-dimensional elliptic problem known as the ``cell problem'' \citep{pavliotis2008}.\footnote{ Note that the ``effective diffusivity'' appearing here should not be confused with the ``effective diffusivity'' in \citet{marshall2006}.}

\begin{table}
  \begin{centering}
  \begin{tabular}{c c c}
    \hline
    Parameter & Symbol & Value(s) \\
    \hline \hline
    Spatial period & $L$ & $100$~km \\
    Maximum vortex speed & $U_\mathrm{TG}$ & $40$~cm~s${}^{-1}$ \\
    Background flow speed & $U_\mathrm{M}$ & $20$~cm~s${}^{-1}$ \\
    Background flow angle & $\phi_\mathrm{M}$ & $30^\circ$ \\
    Small-scale diffusivity & $\kappa$ & $50$~m${}^2$~s${}^{-1}$ \\
    \hline
    Particle integration time step size & $\Delta t$ & $84.3750$~s \\
    Total particle integration time & $T$ & $256$~days \\
    Number of particles & $N$ & $256$ \\
    \hline
    Data sampling interval & $s$ & $3$~hours, $6$~hours, \dots $120$~days\\
    Markov Chain Monte Carlo iterations & $N_{mh}$ & $10^5$ \\
    Number of independent Markov Chains & $M$ & $3$ \\
    \hline
  \end{tabular}\caption{
    Parameters used in the idealised Taylor--Green vortex configuration.}\label{tab:tg}
  \end{centering}
\end{table}

\subsection{Bayesian inference}

We apply Bayesian inference to this problem for the uniform velocity and diffusivity
\begin{subequations}\label{eqn:param_hom}
  \begin{align}
    \vec{U}  & = \vec{U}\left( \svec{\theta} \right) = U \left( 
    \begin{array}{c} 
        \cos \phi \\ \sin \phi 
    \end{array} \right), \\
    \tensor{K} & = \tensor{K}\left( \svec{\theta} \right) = 
      \tensor{R} \left( \phi_\tensor{K} \right) \left(
    \begin{array}{cc}
        \gamma_1 & 0 \\ 0 & \gamma_2 
    \end{array} \right) \tensor{R} \left( \phi_\tensor{K} \right)^\mathrm{T}, \label{eqn:param_hom_diff}
  \end{align}
\end{subequations}
where
\begin{equation}\label{eqn:rotation}
  \tensor{R} \left( \phi_\tensor{K} \right) = \left( 
    \begin{array}{cc}
        \cos \phi_\tensor{K} & -\sin \phi_\tensor{K} \\ 
        \sin \phi_\tensor{K} & \cos \phi_\tensor{K}
    \end{array} \right)
\end{equation}
is a rotation matrix. Thus the parameters to infer are
\begin{equation}
  \svec{\theta} = \left( U, \phi, \gamma_1, \gamma_2, \phi_\tensor{K} \right)^\mathrm{T}.
\end{equation}
The representation \eqref{eqn:param_hom_diff} of the diffusivity $\tensor{K}$ is motivated by its eigendecomposition, and guarantees that it is symmetric positive-definite when $\gamma_1$ and $\gamma_2$ are positive. 

Parameters used in this example are provided in Table~\ref{tab:tg}. The domain size, flow speeds, and small-scale diffusivity are chosen so as to yield an ocean-like regime. Particle trajectory data are generated by solving the SDE \eqref{eqn:ad-diff-TG} for $N$ particles initially located on a uniform square grid in the doubly-periodic domain $\left( x, y \right) \in \left[ -L, L \right]^2$. The SDE is solved numerically using the Euler--Maruyama method with a small timestep size of $\Delta t = 84.3750$~s. For the purposes of the Bayesian inference their positions are sampled with a sampling interval $s=t_j-t_{j-1}$ over a total time of $T = 256$~days.

\subsection{Posterior evaluation}

For the uniform velocity and diffusivity \eqref{eqn:param_hom}, the Fokker--Planck equation can be solved analytically, yielding the Gaussian transition probability density
\begin{equation}
\pi\left(\vec{X}_i \left( t_{j + 1} \right), s  |  \vec{X}_i \left( t_{j} \right)\right) =
\frac{1}{2 \pi \sqrt{\det \Sigma_s}}
      \exp \left( -\frac{1}{2} \left\| \vec{X}_i \left( t_{j + 1} \right) - \vec{m}_s \left( \vec{X}_i \left( t_j \right) \right) \right\|^2_{\Sigma_s^{-1}} \right),
\end{equation}
where
\begin{equation}
    \vec{m}_s \left( \vec{x} \right) = \vec{x} + \vec{U} s,  \quad 
    \Sigma_s = 2 s \tensor{K}
\end{equation}
and, for a suitably sized vector $\vec{v}$,
\begin{equation} \label{eqn:innerproduct}
  \left\| \vec{v} \right\|_{\Sigma_s^{-1}}^2 = \vec{v}^\mathrm{T} \Sigma_s^{-1} \vec{v}.
\end{equation}
This gives an explicit expression for the likelihood \eqref{eqn:likelihood}.

In order to perform the Bayesian inference a prior must be chosen. This is a subjective choice reflecting expected prior knowledge regarding the parameters under consideration (the elements of $\svec{\theta}$) and, except in limiting cases of large data, the result of the inference is dependent upon the choice of prior. The priors for the angles $\phi$ and $\phi_\tensor{K}$ set equal to the uniform distribution, and the priors for the remaining parameters are uniform in the ranges $U \in \left[ 0, 10~\mathrm{m}~\mathrm{s}^{-1} \right]$ and $\gamma_1, \gamma_2 \in \left[ 1~\mathrm{m}^2~\mathrm{s}^{-1}, 10^5~\mathrm{m}^2~\mathrm{s}^{-1} \right]$, and zero elsewhere.

The posterior is evaluated, up to some unknown proportionality constant, as the product of the likelihood and the prior, noting that the proportionality constant is not required by the Metropolis--Hastings algorithm. In total $3$ independent set of $100,000$ samples $\svec{\theta}^{(k)}$ are drawn, and it is verified that the Gelman--Rubin diagnostic criterion (see \ref{app:metropolis_hastings}) is satisfied.

\subsection{Results}

\begin{figure}
    \centering
    \includegraphics[width=0.95\textwidth]{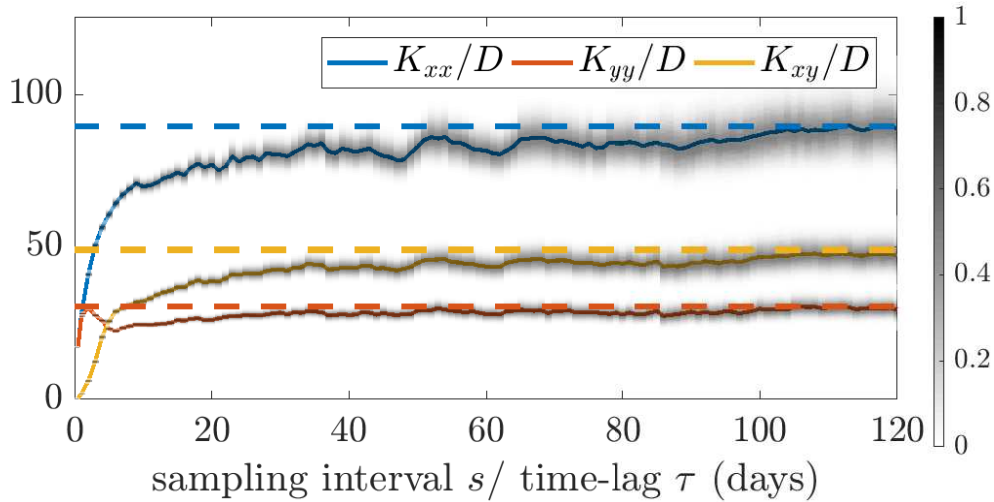}
    \caption{ 
      Results of the Bayesian inference for the diffusivity tensor components for the periodic flow \eqref{eqn:TG} under varying sampling intervals. The posterior distributions are shown with shading, normalised so that the maximum value at each sampling interval is $1$. The posterior means are shown as solid lines. The computed effective diffusivity components of homogenisation theory are shown as dashed lines.}
    \label{fig:tg_w_mean_vs_SIv}
\end{figure}

The posterior mean velocity components (not shown) show little variability with sampling interval and agree excellently with the background flow. The posterior mean diffusivity components are shown in figure \ref{fig:tg_w_mean_vs_SIv}, and show much greater variability. For example, over short time scales the particles experience only local small-scale dynamics, and hence short sampling intervals are associated with low values of inferred diffusivity. The diffusivity components increase with increasing sampling interval and approach a stable value.
As the sampling interval increases, the number of particle positions used in the inference decreases (since the same length of particle trajectory is considered in all cases). As a result, the uncertainty of the inference increases, leading to a widening of the posterior distribution.

For reference the effective diffusivity $\tensor{K}$ of homogenisation theory is computed by solving the elliptic ``cell problem''  \citep[e.g. Eq.\ (2.2) of ][]{cotter2009}. The equations are solved using degree-one continuous Lagrange finite elements using the FEniCS system \citep{logg2012, alnaes2015} version 2018.1.0. A finite element mesh is formed via a $512 \times 512$ structured and uniform square mesh, with each square divided along the lower-left to upper-right diagonal to form a triangle mesh. The results are shown with dashed lines in figure \ref{fig:tg_w_mean_vs_SIv}. The larger sampling interval posterior mean diffusivity components, obtained using Bayesian inference, agree well with the computed effective diffusivity.

\section{Quasigeostrophic double gyre}\label{sect:qg}
The Bayesian inference machinery, illustrated in the preceding section for a highly idealised example, is now applied in a more oceanographically relevant context by considering Lagrangian particle trajectories in a quasigeostrophic double-gyre calculation.

\subsection{Numerical model}

The three-layer quasigeostrophic double gyre configuration of \citet{maddison2015} is considered \citep[see also][]{berloff2007,karabasov2009,marshall2012}. The three-layer quasigeostrophic equations \citep[see][section 3.1]{maddison2015} are discretised using finite differencing, with a mesh with $513 \times 513$ nodes uniformly spaced on a square grid, in a $3840\mathrm{~km} \times 3840\mathrm{~km}$ square horizontal domain. The advection term in the quasigeostrophic potential vorticity equation is discretised using the \citet{arakawa1966} Jacobian, and Laplace operators are discretised using second order centered differencing. The elliptic problem for potential vorticity inversion is solved via projection onto discrete baroclinic modes, and the resulting Poisson or modified Helmholtz problems are solved using a Fast Poisson Solver \citep[e.g.][section 5.5]{strang1986}, with the decoupled tri-diagonal systems arrived at using a Discrete Sine Transform using FFTPACK 5.1. The system is integrated in time using a third-order Adams--Bashforth scheme with uniform timestep $\Delta t_{QG} = 1800$~s.
Physical parameters are as in Table 1 of \citet{maddison2015}.

\subsection{Particle advection}

Particles are advected using the geometric integration approach described in \citet{ham2006a} and \citet{ham2006b}. A piecewise linear streamfunction is constructed from the finite-difference grid point values by dividing each square cell corner-to-corner to yield four isosceles triangles, bi-linearly interpolating to yield a value at the centre vertex, and then linearly interpolating within the triangles. The time-dependent streamfunction is further linearly interpolated in time. Initial starting cells are determined using a quad-tree based search \citep{samet1984} using code derived from libsupermesh \citep{panourgias2016}, after which they are advected along contours of the discrete streamfunction. Note that care needs to be taken to ensure that the particle advection -- which is a two-dimensional computational geometry problem -- is solved in a precision-robust manner. A useful property of the particle advection scheme is that, given a streamfunction which is constant on the boundary, particles are guaranteed to never leave the bounding domain \citep[see][]{ham2006a}. Hence the particle advection scheme requires no further consideration of boundary condition.

\begin{figure}
    \centering
    \includegraphics[width=\textwidth]{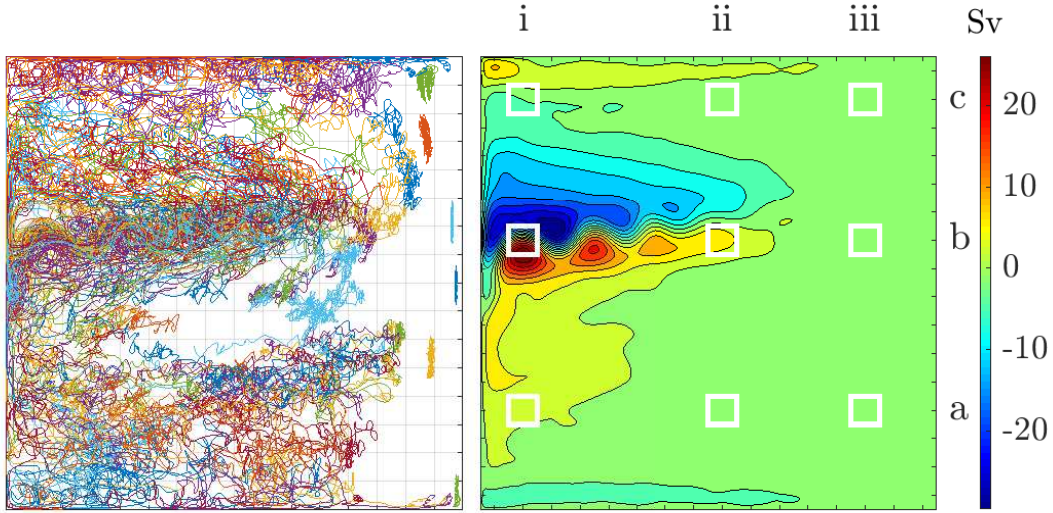}    
    \caption{
      Left panel: $10$-year trajectories for $60$ arbitrarily selected particles in the middle layer of the quasigeostrophic double-gyre system. The division of the domain into a $16 \times 16$ array of square elements is shown in grey.
      Right panel:  $10$-year time-averaged streamfunction, multiplied by the layer thickness, in the middle layer. Selected cells of the $16 \times 16$ array referred to in the main text are highlighted in white and labelled by a letter/numeral coordinate.}
    \label{fig:QGM2_DStemp_TrajVisWithRect_Idx16_NParts60}
\end{figure}

We consider only particle advection, with no explicit small-scale diffusivity, within the middle layer of the model. This layer experiences no direct wind forcing or bottom linear drag. After a $100$~year spinup\footnote{Julian years are used throughout.} $625$ particles are distributed uniformly across the square domain. This number is chosen so as to resemble the typical number of ARGO drifters available in the North Atlantic \citep{argo2000}. The particles are then advected for a further $10$~years, and their positions are recorded daily. The resulting trajectories for $60$ arbitrarily selected particles are shown in figure \ref{fig:QGM2_DStemp_TrajVisWithRect_Idx16_NParts60}.

\subsection{Bayesian inference}

The domain is partitioned into a $16 \times 16$ array of square cells with $240$~km side lengths. 
Within each cell the velocity is represented as a linearly varying non-divergent field, and the diffusivity as a constant symmetric positive definite tensor,
\begin{subequations} \label{eqn:param_linear}
  \begin{align}
    \vec{U} \left( \vec{x} \right) & = \vec{U}\left( \vec{x} ; \svec{\theta} \right) = 
      \tensor{A} \left( \vec{x} - \vec{x}_0 \right) + \vec{u}_0, \\
    \tensor{K} & = \tensor{K} \left( \svec{\theta} \right) = 
      \tensor{R} \left( \phi_\tensor{K} \right) \left( \begin{array}{cc} \gamma_1 & 0 \\ 0 & \gamma_2 \end{array} \right) \tensor{R} \left( \phi_\tensor{K} \right)^\mathrm{T},
  \end{align}
\end{subequations}
where
\begin{subequations} \label{eqn:param_linear_sym}
  \begin{align}
    \vec{u}_0 & = \vec{u}_0 \left( \svec{\theta} \right) =  U_0 \left( \begin{array}{c} \cos \phi_0 \\ \sin \phi_0 \end{array} \right), \\
    \tensor{A} & = \tensor{A} \left( \svec{\theta} \right) = R \left( \phi_\tensor{A} \right) \left( \begin{array}{cc} 0 & \nu_2 + \nu_1 \\ \nu_2 - \nu_1 & 0 \end{array} \right) R \left( \phi_\tensor{A} \right)^\mathrm{T}.
  \end{align}
\end{subequations}
Here $\tensor{R}$ is a rotation matrix as in \eqref{eqn:rotation} and $\vec{x}_0$ is the centre of the cell.
Thus the parameters to infer in each cell are
\begin{equation}
  \svec{\theta} = \left(
    U_0,
    \phi_0,
    \nu_1,
    \nu_2,
    \phi_\tensor{A},
    \gamma_1,
    \gamma_2,
    \phi_\tensor{K} \right)^\mathrm{T}.
\end{equation}
The linear velocity field introduces additional degrees of freedom compared with the uniform velocity field used in section \ref{sect:idealised}. It is motivated by the large shears that are present in the jet region of the simulation.

\subsection{Posterior evaluation}

The Fokker--Planck equation can be solved analytically for the velocity and diffusivity \eqref{eqn:param_linear}, yielding the Gaussian transition probability density
\begin{equation}
    \pi\left(\vec{X}_i \left( t_{j + 1} \right), s  |  \vec{X}_i \left( t_{j} \right)\right) = \frac{1}{2 \pi \sqrt{\det \left( \Sigma_s \right)}}
      \exp \left( -\frac{1}{2} \left\| \vec{X}_i \left( t_{j + 1} \right) - \vec{m}_s \left( \vec{X}_i \left( t_j \right) \right) \right\|^2_{ \Sigma_s^{-1}} \right),
\end{equation}
where
\begin{subequations}
  \begin{align}
    \vec{m}_s \left( \vec{x} \right) &= \vec{m}_s \left( \vec{x}; \svec{\theta} \right) = e^{\tensor{A} s} \vec{x}  + \int_0^s e^{\tensor{A} t} \mathrm{d}t \left( \vec{u}_0 - \tensor{A} \vec{x}_0 \right), \\
    \Sigma_s & = \Sigma_s \left( \svec{\theta} \right) = 2 \int_0^s e^{\tensor{A} t} \tensor{K} e^{ \tensor{A}^\mathrm{T} t} \mathrm{d}t,
  \end{align}
\end{subequations}
and $\left\| \vec{v} \right\|_{\Sigma_s^{-1}}$ is defined in \eqref{eqn:innerproduct}. This gives an explicit expression for the likelihood \eqref{eqn:likelihood}.

We take again simple uniform priors for $p(\svec{\theta})$: the angles $\phi_0$, $\phi_\tensor{A}$, and $\phi_\tensor{K}$ are uniform, and remaining parameters are uniformly distributed in the ranges $U_0 \in \left[ 0, 10~\mathrm{m}~\mathrm{s}^{-1} \right]$,  $\nu_1, \nu_2 \in \left[ -10^{-5}~\mathrm{s}^{-1}, 10^{-5}~\mathrm{s}^{-1} \right]$, $\gamma_1, \gamma_2 \in \left[ 1~\mathrm{m}^2~\mathrm{s}^{-1}, 10^5~\mathrm{m}^2~\mathrm{s}^{-1} \right]$ and are zero elsewhere. It has been verified that the results would be unaffected if these ranges were extended.

The posterior is evaluated, up to some unknown proportionality constant, as the product of the likelihood and the prior. In total $10$ independent chains of $4 \times 10^5$ samples $\svec{\theta}^{(k)}$ are then drawn using the Metropolis--Hastings algorithm and the \citet{gelman1992} diagnostic (also in \ref{app:metropolis_hastings} and section 11.4 of \cite{gelman2013}) to test convergence. This process is performed separately for each cell of the  $16 \times 16$ array covering the model domain. We consider the sampling intervals $s=1,\,  2, \, 4,\, 8, 16,\, 32,\, 48, \, \cdots,\,  128$ days.  The samples of each of the independent chains are combined to approximate the posterior distribution.

\subsection{Results}
\begin{figure}
    \centering
    \includegraphics[width=\textwidth]{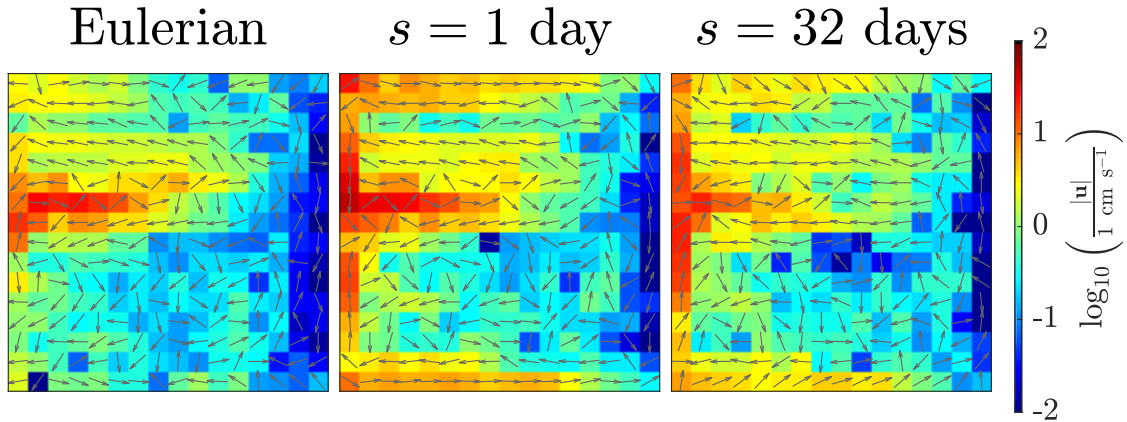}    
    \caption{
      Left: $10$-year Eulerian mean velocity in the middle layer at the cell centres.
      Middle and right:  MAP estimate for the cell-centre middle layer velocity using particle positions observed at sampling intervals of $s = 1$~day and $s = 32$~days.
      The magnitude of the mean velocity is shown using a logarithmic colour scale, and the velocity direction is indicated by equal-length arrows.
      }
    \label{fig:QGM2_DStemp_UVEulerArrow_diffSIv_Idx16_LocLinINC}
\end{figure} 

Figure \ref{fig:QGM2_DStemp_UVEulerArrow_diffSIv_Idx16_LocLinINC} shows the maximum a posteriori estimate (MAP) for the middle-layer  velocity field, together with the Eulerian mean flow computed over the $10$-year data collection window. The MAP estimate of $\svec{\theta}$ is the maximiser for the posterior $p \left( \svec{\theta} | R \right)$ and indicates the most likely combination of mean flow and diffusivity fields to recover the trajectory data. In all cases described here the MAP estimate is approximated by the sample $\svec{\theta}^{(k)}$ that maximises the posterior $p(\svec{\theta}^{(k)}|R)$ over all MCMC steps $k$ and over all chains. For a short sampling interval $s = 1$~day, the MAP flow velocity is comparable to the Eulerian mean velocity.

\begin{figure}
    \centering
    \includegraphics[width=\textwidth]{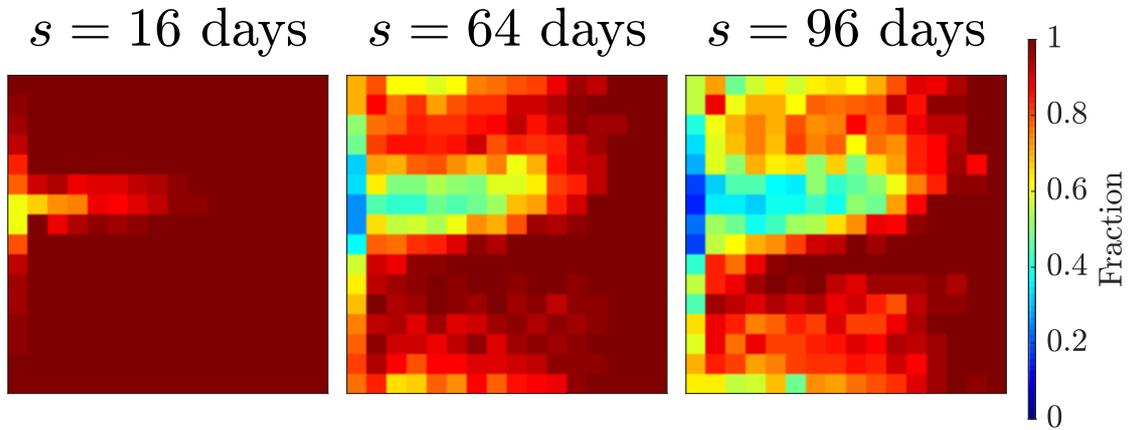}
    \caption{
      Fractions of particles found in their origin cell, or in the eight cells surrounding the origin cell, at the end of the sampling interval. For sampling intervals shorter than $4$ days all particles remain in this neighbourhood. }
    \label{fig:QGM2_DStemp_RemainRatio_diffSIv_Idx16}
\end{figure} 

Figure \ref{fig:QGM2_DStemp_RemainRatio_diffSIv_Idx16} shows the fraction of particles which are found in their cell of origin or in one of the eight surrounding cells at the end of sampling interval (regardless of the intermediate trajectory). This provides an indication of the validity of the locality assumption inherent in the local inference approach. For short sampling intervals this fraction is high, but as expected it drops as the sampling interval increases; in particular, it drops to very low values in the jet and on the western boundary. There is therefore potential misattribution of the spatial location of flow properties in these regions. This is a significant issue on the western boundary, where particles flow rapidly from the boundary into the jet, and rapidly change direction from a northward or southward flow, to an eastward flow.

\begin{figure}
    \centering
    \includegraphics[width=\textwidth]{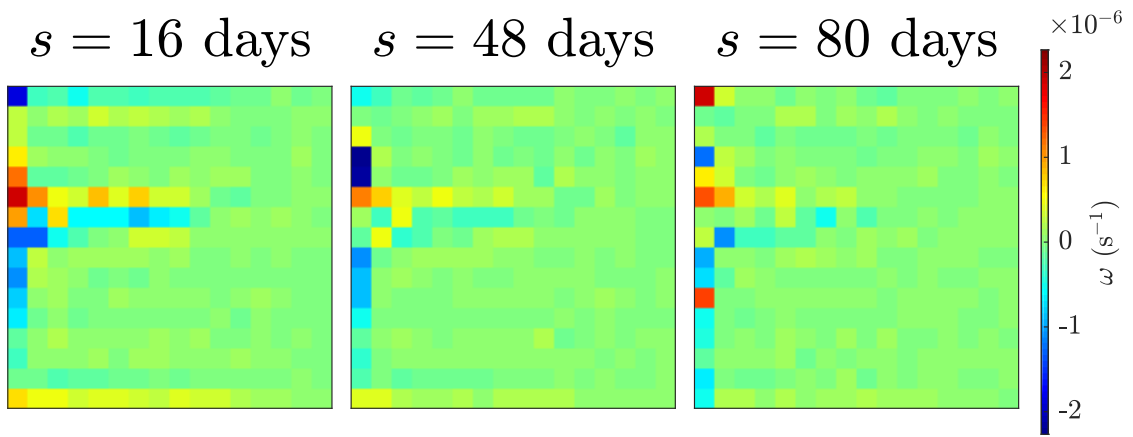}    
    \caption{
      MAP estimate of the  vorticity $\omega= \partial_x v  - \partial_y u$ field at different sampling intervals $s$.
      }
    \label{fig:QGM2_DStemp_Vorticity_diffSIv_Idx16_LocLinINC}
\end{figure} 

At short sampling intervals $(\lesssim 16~\mathrm{days})$, strong shears are inferred along the jet and on the northern, western, and southern boundaries. This is indicated by the large local vorticity $\omega= \partial_x v  - \partial_y u$, corresponding to the off-diagonal elements of $\tensor{A}-\tensor{A}^\mathrm{T}$, shown in figure \ref{fig:QGM2_DStemp_Vorticity_diffSIv_Idx16_LocLinINC}. The inferred diffusivity in these areas is significantly reduced (not shown) when the spatial gradients of the mean flow are resolved, by permitting a non-zero linear shear. For the large sampling intervals the inferred shear tensor is smaller, as may be expected for a Lagrangian average of the flow over these time scales. Hence for the large sampling intervals the inferred diffusivity is largely unaffected by the inclusion of shear in $\vec{U}(\vec{x} ; \svec{\theta})$, and an inference with a locally constant velocity would yield similar results.

\begin{figure}
    \centering
    \includegraphics[width=\textwidth]{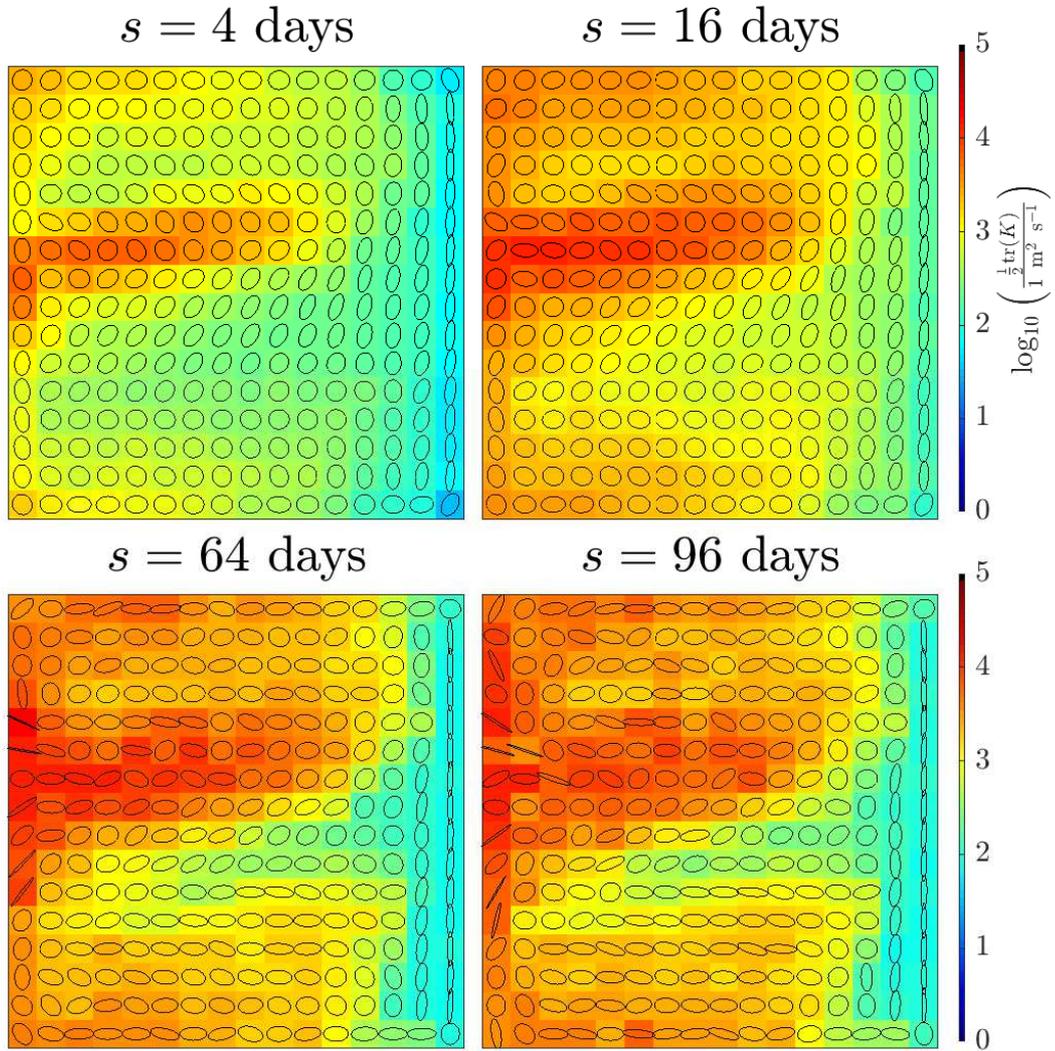}
    \caption{MAP estimate of the middle layer diffusivity field at different sampling intervals $s$. The (logarithmic) colour scale gives the half trace of the diffusivity tensor $\tensor{K}$, which is also the arithmetic mean of the eigenvalues, to characterise the magnitude of diffusivity. The ellipses visualise the directions and the relative magnitude of the two eigenvectors of the diffusivity tensor in each cell.}
    \label{fig:QGM2_DStemp_DiffusivityEllipses_diffSIv_Idx16_LocLinINC}
\end{figure}

Figure \ref{fig:QGM2_DStemp_DiffusivityEllipses_diffSIv_Idx16_LocLinINC} visualises the MAP estimate for the middle layer diffusivity tensor for differing sampling intervals. The ``diffusivity ellipses'' in figure \ref{fig:QGM2_DStemp_DiffusivityEllipses_diffSIv_Idx16_LocLinINC} outline the orientations of contours of a passive tracer if it undergoes pure diffusion with a Dirac-delta initial profile, characterising the directions of the anisotropy of the eddy diffusion tensor. The diffusivity magnitude, defined as the half trace of the diffusivity tensor, is visualised using the colour scale. The inferred diffusivity is largest in the jet region, and strengthens with increasing sampling interval. There is a region of very weak inferred diffusivity in the eastern part of the southern half of the domain. At large sampling interval the anisotropic diffusion has a preferential east-west orientation in the gyres and the core of the jet. Near the western boundary the anisotropic diffusivity is tilted towards the direction of the jet -- this is attributed to non-local effects, as particles are rapidly transported into the jet from this region.

\begin{figure}
    \centering
    \includegraphics[width=\textwidth]{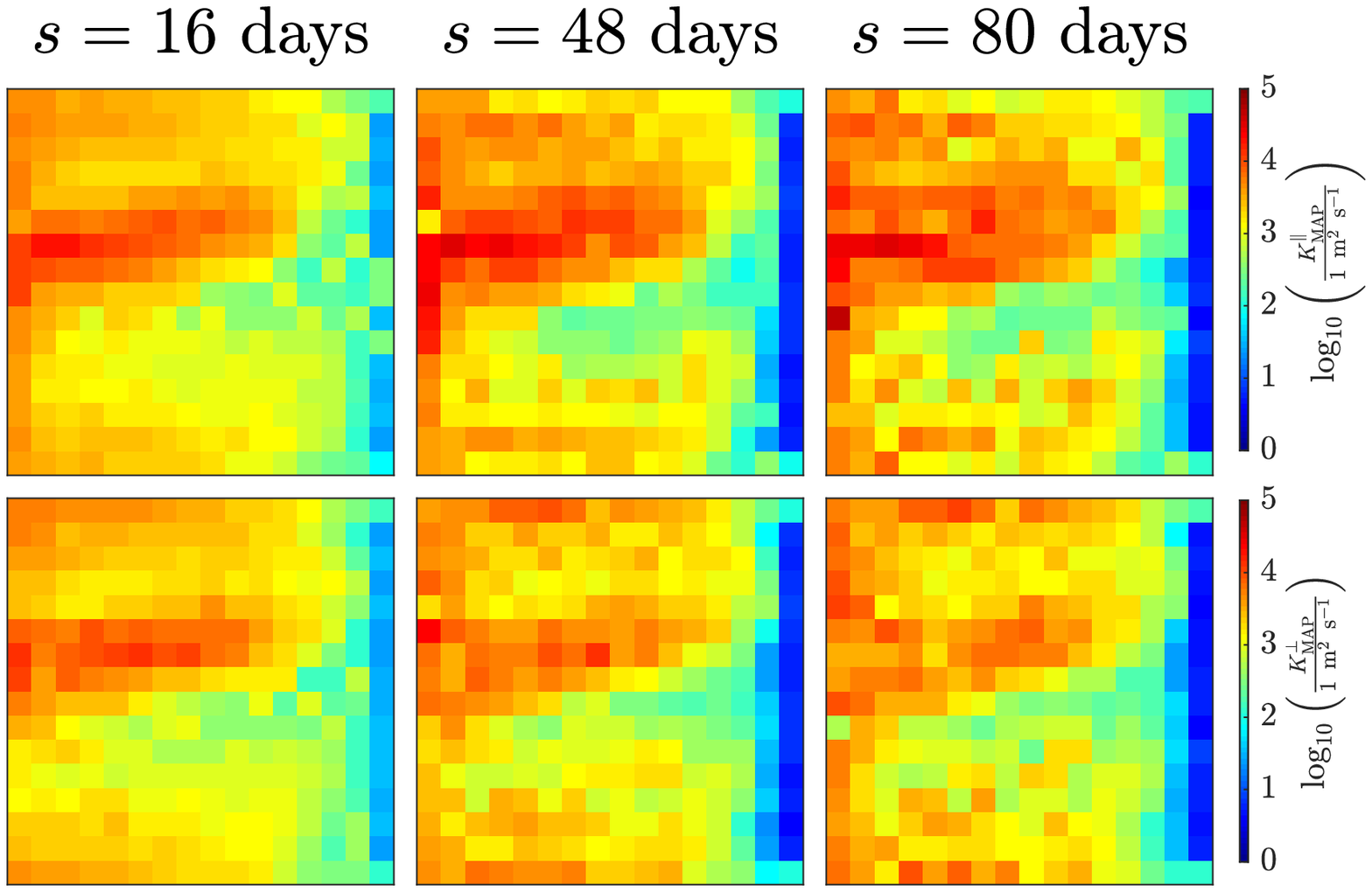}
    \caption{
       MAP estimate for the diffusivity components in the middle layer with different sampling intervals. 
      Upper panels:  along-stream diffusivity; lower panels: cross-stream diffusivity. 
      A logarithmic colour scale is used.}
    \label{fig:QGM2_DStemp_KBayes_diffSIv_Idx16_LocLinINC}
\end{figure}

\begin{figure}
    \centering
    \includegraphics[width=\textwidth]{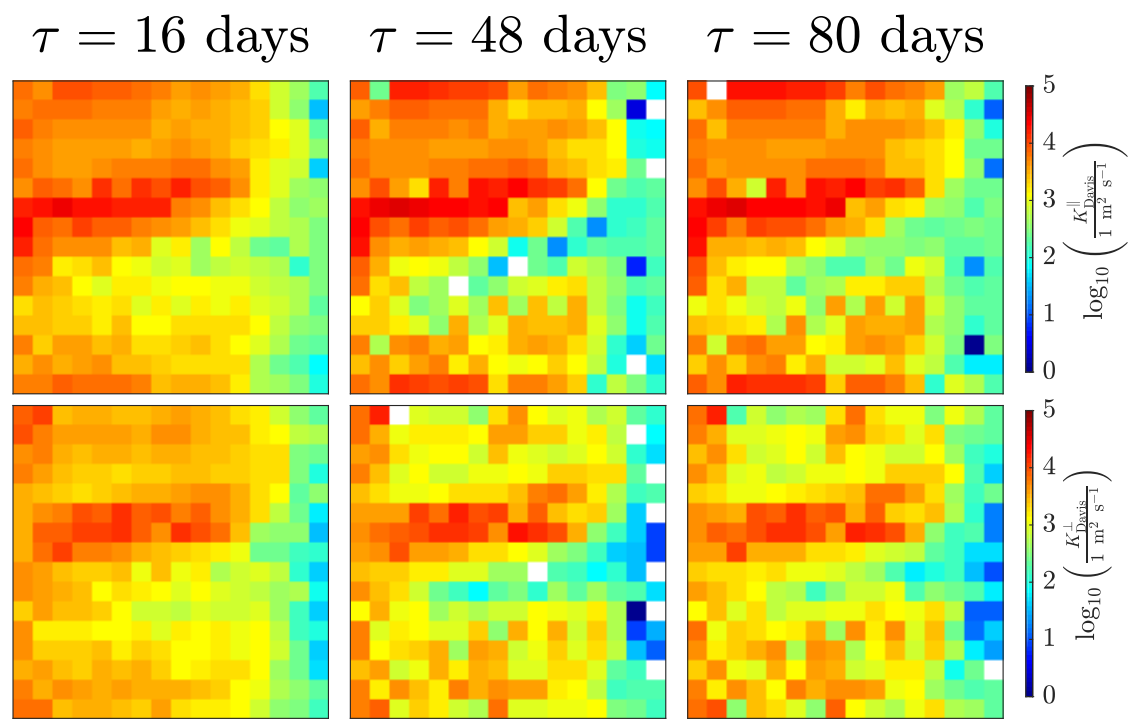}
    \caption{
       \citet{davis1987} diffusivity components in the middle layer with different time lags. 
      Upper panel: along-stream diffusivity; lower panel:  cross-stream diffusivity. 
      A logarithmic colour scale is used. Missing data, shown in white, correspond to negative values of the \citet{davis1987} diffusivity components.}
    \label{fig:QGM2_DStemp_KDavis_diffSIv_Idx16_LocLinINC}
\end{figure} 

\begin{figure}
    \centering
    \includegraphics[width=\textwidth]{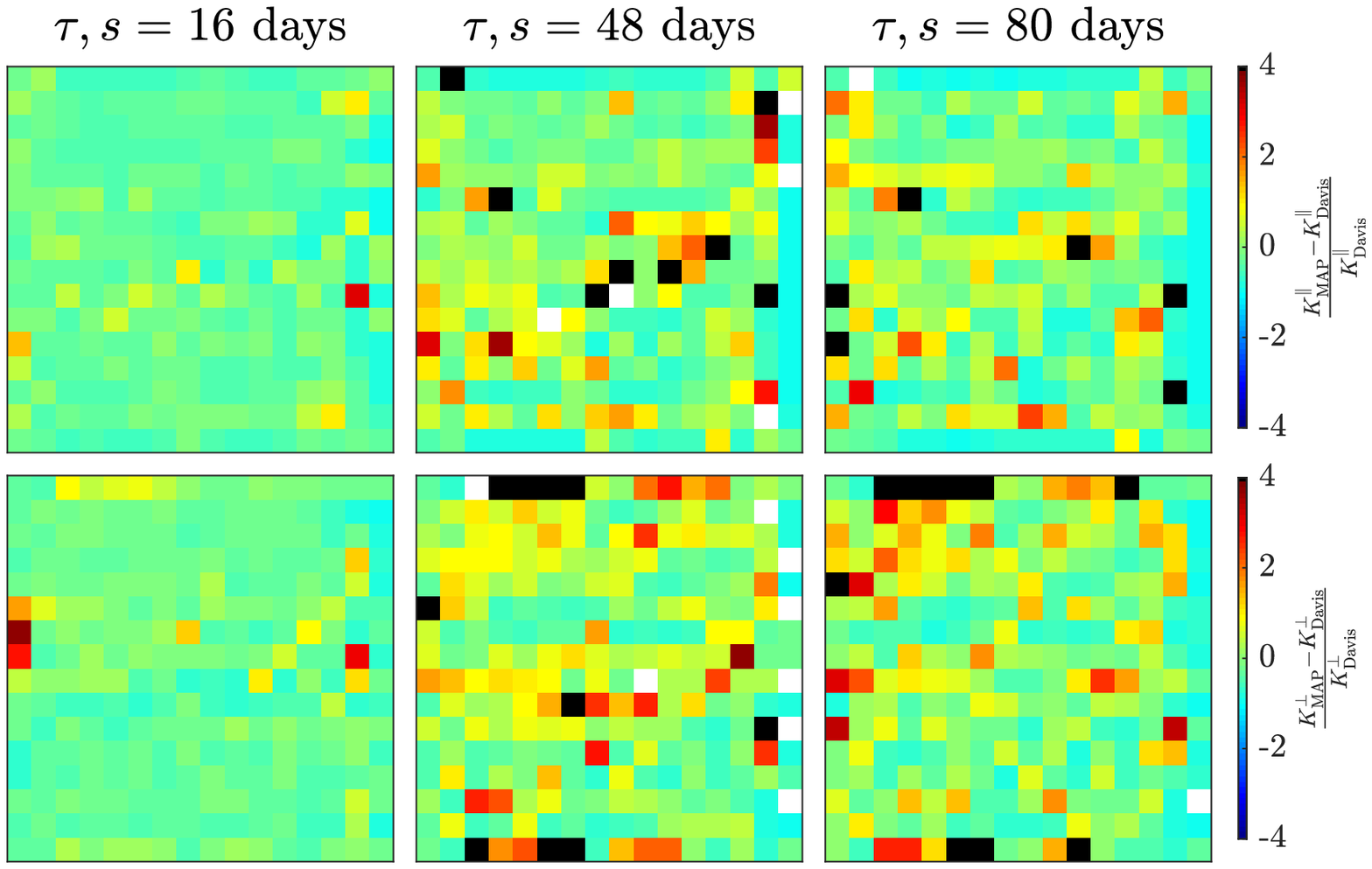}
    \caption{
      Relative difference between \citet{davis1987} and MAP estimates of middle layer diffusivity components with different sampling intervals/time lags. 
      Upper panel: along-stream diffusivity; lower panel:  cross-stream diffusivity. 
      Missing data, shown in white, correspond to negative values of the \citet{davis1987} diffusivity. 
      Data with values exceeding the range of visualisations are shown in black.}
    \label{fig:QGM2_DStemp_Kratio_diffSIv_Idx16_LocLinINC}
\end{figure} 

The Metropolis--Hastings algorithm samples the joint posterior distribution of the velocity and diffusivity and so makes it possible to infer quantities that depend on both fields.
In particular, we can construct distributions for the cross-stream and along-stream diffusivity components $K^\bot$ and $K^\parallel$ by projecting for each sample $k$, the sample diffusivity $\tensor{K}^{(k)}$, in directions perpendicular to and parallel to the sample velocity $\vec{U}^{(k)}$.
The resulting MAP estimates are shown in figure \ref{fig:QGM2_DStemp_KBayes_diffSIv_Idx16_LocLinINC}. For comparison, the cross-stream and along-stream \citet{davis1987} diffusivity, defined with respect to the $10$~year Eulerian mean flow at the cell centre, are shown in figure \ref{fig:QGM2_DStemp_KDavis_diffSIv_Idx16_LocLinINC}.

The two diagnostic approaches generally agree well in order of magnitude and spatial structure, with increased diffusivity in the region of the jet and reduced diffusivity on the eastern boundary and in the region south of the jet, as indicated by their relative differences in figure \ref{fig:QGM2_DStemp_Kratio_diffSIv_Idx16_LocLinINC}. There is some disagreement in detail, for example near the northern and southern boundaries. Note that the \citet{davis1987} diffusivity as computed here is not a symmetric positive definite (or even symmetric) quantity in general, leading to some regions of missing data indicated in white in figures \ref{fig:QGM2_DStemp_KDavis_diffSIv_Idx16_LocLinINC} and \ref{fig:QGM2_DStemp_Kratio_diffSIv_Idx16_LocLinINC}.

\begin{figure}
    \centering
    \includegraphics[width=\textwidth]{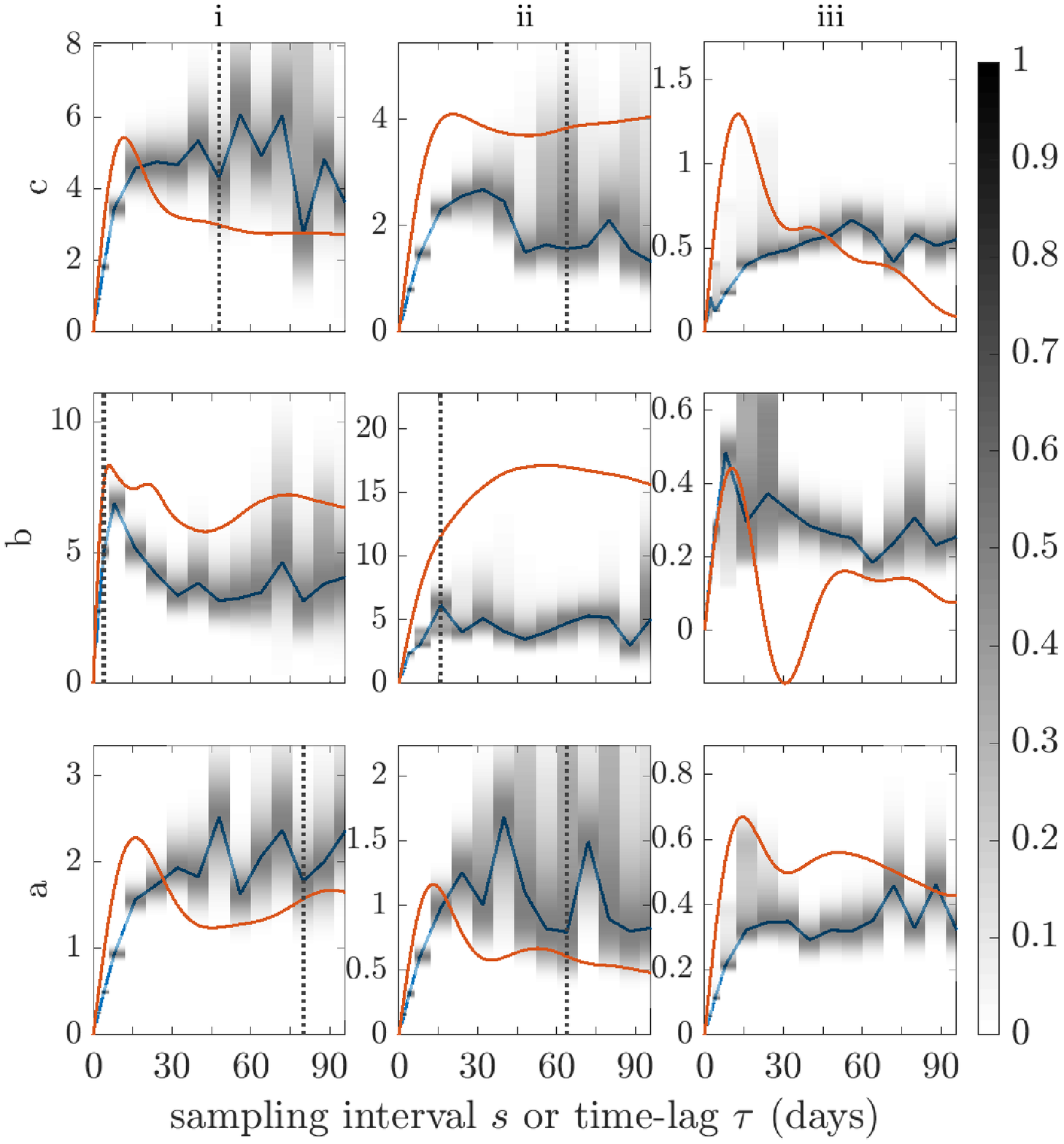}
    \caption{Cross-stream diffusivity $K^{\perp}$ (in $1000 ~\mathrm{m}^{2}~\mathrm{s}^{-1}$) in the middle layer against sampling interval $s$ or time-lag $\tau$ in selected cells, labeled on the top of each column and left of each row (see figure \ref{fig:QGM2_DStemp_TrajVisWithRect_Idx16_NParts60}). The blue lines are the MAP estimates of the Bayesian inference; the red lines correspond to the \citet{davis1987} diffusivity.
    The grey shading shows the marginal posterior density for $K^{\perp}$, normalised by its maximum values for each $s$.  The dash-dot vertical lines indicate the time taken for $10$ percent of particles to exit the origin and its neighbouring $8$ cells. Note that the vertical line is not shown if this time is beyond $128$ days. }
    \label{fig:QGM2_DStemp_Npart625_Kperp_diffSIv_Idx16_LocLinINC}
\end{figure}

\begin{figure}
    \centering
    \includegraphics[width=\textwidth]{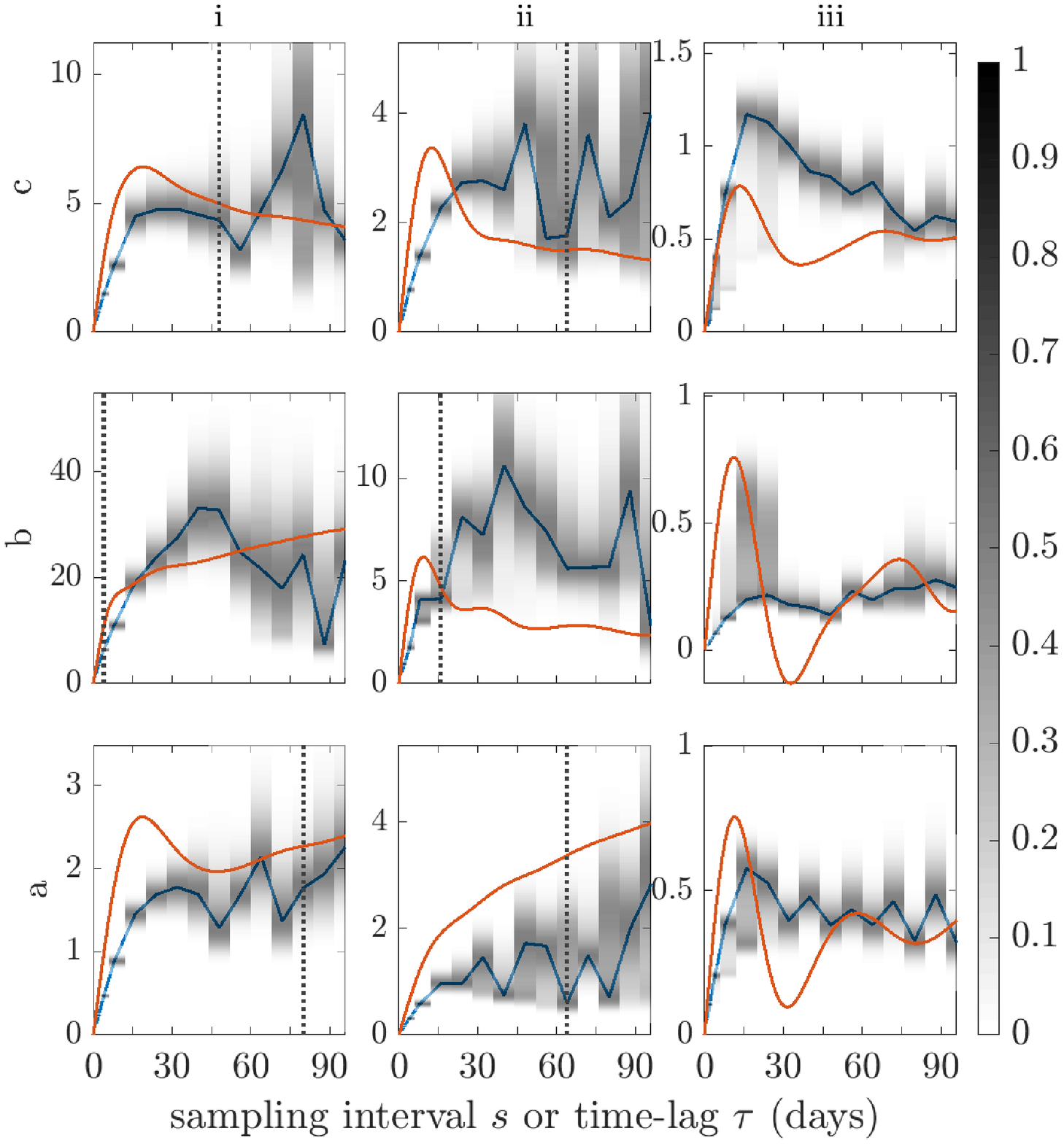}
    \caption{Same as figure \ref{fig:QGM2_DStemp_Npart625_Kperp_diffSIv_Idx16_LocLinINC} but for the along-stream diffusivity $K^{\parallel}$ in the middle layer.}
    \label{fig:QGM2_DStemp_Npart625_Kstrm_diffSIv_Idx16_LocLinINC}
\end{figure}

To analyse our results in more detail, we now focus on the 9 cells highlighted in figure 
\ref{fig:QGM2_DStemp_TrajVisWithRect_Idx16_NParts60} and labelled (i)--(iii) with increasing $x$ coordinate, and (a)--(c) with increasing $y$ coordinate.
Figures \ref{fig:QGM2_DStemp_Npart625_Kperp_diffSIv_Idx16_LocLinINC} and \ref{fig:QGM2_DStemp_Npart625_Kstrm_diffSIv_Idx16_LocLinINC} show the MAP of the middle layer cross-stream and along-stream diffusivity in these cells as functions of the sampling interval $s$. The \citet{davis1987} diffusivity is shown for comparison; the time lag $\tau$ and sampling interval $s$ are shown on a common scale even though the two parameters are not strictly comparable.
The MAP diffusivities do demonstrate a degree of convergence at larger sampling intervals, and agree in order of magnitude, at larger sampling intervals, with the large time-lag \citet{davis1987} diffusivity. The MAP diffusivities are never negative, as a consequence of the choice of prior, and while some variation is observed with sampling interval, the Bayesian diffusivity estimates are generally more stable in magnitude than the \citet{davis1987} diffusivity values.

One of the advantages of the Bayesian approach is that it provides a probability distribution, rather than single estimates for $\svec{U}$ and $\tensor{K}$, and hence allows for a quantification of the uncertainty. This is illustrated in Figures \ref{fig:QGM2_DStemp_Npart625_Kperp_diffSIv_Idx16_LocLinINC} and \ref{fig:QGM2_DStemp_Npart625_Kstrm_diffSIv_Idx16_LocLinINC} which also show the (marginal) posterior probability density for the two diffusivity components at each sampling interval $s$. The probability densities are shown as shading and normalised by their maximum value at each value of $s$. 
Broadly speaking, the figures suggest that the range of plausible values is reasonably well constrained, with low probabilities for values more than a factor of, say 2, away from the MAP. Nevertheless, relatively long tails of the posterior distribution indicate that there is a significant probability of diffusivities of much larger magnitude that the MAP values. There are cases of multi-modality, for example in the lower right panel of figure \ref{fig:QGM2_DStemp_Npart625_Kperp_diffSIv_Idx16_LocLinINC} and figure \ref{fig:QGM2_DStemp_Npart625_Kstrm_diffSIv_Idx16_LocLinINC}, with in this cases a MAP value which switches between the two local maxima. 
We attribute this to weakness of the flow in these regions which leads to an ambiguity in the flow direction and hence in the decomposition between along-stream and cross-stream diffusivity.


\section{Conclusions and future work}\label{sect:conclusions}

This article introduces the application of Bayesian inference to the diagnosis of eddy diffusivities from Lagrangian trajectory data. Assuming that the trajectories are governed by a stochastic differential equation involving a number of parameters, 
the Bayesian inference machinery provides an objective way of incorporating all available data so as to yield a full multidimensional posterior probability distribution for the parameters, which quantifies their plausibility. We utilise this to estimate both an anisotropic diffusivity tensor and a linearly varying non-divergent velocity, and to quantify the uncertainty of the estimates. 

Note that the posterior distribution has a very specific interpretation: it is a probability density for the parameters, assuming a perfect model, and given the data and prior information. The posterior can exhibit spread due to lack of data, as weighted against the prior, but not due to error in the underlying model. Further, while we may anticipate convergence with increasing particle number or sampling interval, such limits may in practical oceanographic cases not be achievable.

An idealised experiment, consisting of Taylor--Green vortices embedded within a background flow, is considered. Here, with sufficiently long sampling intervals, the inferred diffusivity components agree well with the predictions from homogenisation theory. In a more complex quasigeostrophic model of a three-layer oceanic gyre system, a ``local'' approach is applied to infer the middle layer mean flows and diffusivities independently in each of $16 \times 16$ cells partitioning the domain. The results show that the data of 625 trajectories over 10~years constrain the diffusivity within a factor of about 2 in most of the domain. The values found become relatively insensitive to the sampling time when this exceeds 30 days or so and are roughly comparable to the \citet{davis1987} diffusivity.

We emphasise that the Bayesian approach provides a general framework for the inference of diffusivity which extends well beyond the simple implementation presented in this paper. A crucial limitation of this implementation is the assumption of locality, which supposes that particles observed from a given cell are advected by the same flow velocity and experience the same diffusivity over the entire sampling interval. Even with the relatively large size of cells considered here (240 km), this assumption is problematic, especially near the western boundary and in the region of the separated jet, where the trajectories of many particles straddle several cells. This limitation is not inherent to the Bayesian framework and can in principle be overcome by considering a discretisation of the velocity and diffusivity over the entire domain, and inferring all associated degrees of freedom simultaneously. Two challenges need to be addressed in this more general case: first, the MCMC sampling of the posterior distribution needs to be performed over a space of much higher dimension; second, the transition probability, which solves a Fokker--Planck (i.e., advection--diffusion) equation with spatially varying velocity and diffusivity, cannot be evaluated in closed form. The first challenge is not necessarily a major one: theoretical results \citep{roberts1997} suggests that the complexity of the simultaneous sampling of all the parameters need not be markedly different from that of the combined sampling of the (independent) parameters associated with a single cell. 
The second challenge requires efficient methods to compute, likely in an approximated form, the transition probability from the Fokker--Planck equation. This is the subject of ongoing work. 

In addition to offering a systematic method to make best use of all available data to estimate diffusivities, the Bayesian approach has the advantage of providing a quantification of the uncertainty of these estimates by means of a complete probability density function. This is important when the amount of data is limited, e.g.\ for estimates based on real drifters as opposed to simulated trajectories, and could be used prior to measurement campaigns to help assessing how many drifters are needed. 
Beyond this, we also note that a Bayesian approach can be employed for model selection and determine whether stochastic differential equations more sophisticated than \eqref{eqn:sde} \citep[e.g., as in][]{berloff2002b} are necessary to explain observed trajectories. This is another direction of future work. 



\paragraph{Acknowledgments}
YKY was supported by the Principal's Career Development PhD Scholarships and Edinburgh Global Scholarships. YKY acknowledges advice from Alexa Griesel regarding the \citet{davis1987} diffusivity. The authors are grateful to Luis Zavala Sans\'on for useful comments on the manuscript. 

\appendix

\section{Metropolis--Hastings algorithm} \label{app:metropolis_hastings}
\subsection{Algorithm outline}
The Metropolis--Hastings Algorithm \citep[e.g. section 11.2 of][]{gelman2013} is outlined as follows.
\begin{enumerate}
    \item Set $k=0$. Choose a proposal density $P(\cdot|\cdot)$ and take an initial value for the parameter $\svec{\theta}^{(0)}$.
    \item Iterate: 
    \begin{enumerate}
        \item randomly draw a candidate parameter $\svec{\vartheta}$ with probability $P(\svec{\vartheta} | \svec{\theta}^{(k)})$,
        \item compute $P(\svec{\vartheta} | \svec{\theta}^{(k)})$ and $P(\svec{\theta}^{(k)} | \svec{\vartheta})$, 
        \item compute $p ( \svec{\theta}^{(k)} | R )$ and $p \left( \svec{\vartheta} | R \right)$ (up to an irrelevant proportionality constant) from Bayes' formula \eqref{eqn:bayes}, using the fields $(\vec{u}(\vec{x}; {\svec{\theta}^{(k)}}),\tensor{K}(\vec{x}; {\svec{\theta}^{(k)}}))$ or $(\vec{u}(\vec{x}; \svec{\vartheta}),\tensor{K}(\vec{x}; \svec{\vartheta}))$ for the transition probability $\pi$ ,
        \item let 
        \begin{equation}
        \svec{\theta}^{(k+1)} = \begin{cases}
        \svec{\vartheta} & \textrm{with probability} \ \alpha, \\
        \svec{\theta}^{(k)} & \textrm{with probability}\  1-\alpha,
        \end{cases}
        \end{equation}
        where
        \begin{equation} \label{eqt:AccRatio_MCMC}
            \alpha = \min\left(1,\frac{p ( \svec{\vartheta}|R )}{ p ( \svec{\theta}^{(k)} | R )  } \frac{P(\svec{\theta}^{(k)}|\svec{\vartheta})}{P(\svec{\vartheta}|\svec{\theta}^{(k)})} \right), 
        \end{equation}
        \item increment $k \mapsto k+1$.
    \end{enumerate}
\end{enumerate}

The proposal density  $P(\svec{\vartheta} | \svec{\theta}^{(k)})$  should be easy to compute.
In this paper, we take it such that all the components of $\svec{\vartheta}$ but one are the same as the components of $\svec{\theta}^{(k)}$ -- a technique known as the ``Gibbs sampler'' \citep{geman1984}. Specifically, we take it as the Gaussian
\begin{equation}
    P(\svec{\vartheta} | \svec{\theta}^{(k)}) = \frac{1}{J} \sum_{j=1}^J \frac{1}{ \sqrt{2\pi V_j} } \exp{\left( - \frac{\left( \vartheta_j - \theta^{(k)}_{j} \right)^2}{2 {V_j}} \right)} 
    \prod_{i \ne j}\delta \left( \theta^{(k)}_{i} - \vartheta_i \right),
\end{equation} 
where $j=1,\cdots,J$ labels the components of $\svec{\theta}$ and the variances $V_j$ are tuned for efficient sampling (see below).  Note that  $P(\svec{\vartheta} | \svec{\theta}^{(k)}) = P( \svec{\theta}^{(k)} | \svec{\vartheta})$, which simplifies the form of $\alpha$ in \eqref{eqt:AccRatio_MCMC} and eliminates the need for step 2(b). 





It should be noted that it is only the distribution of $\svec{\theta}^{(k)}$ (the stationary distribution) that converges to the target posterior $p(\svec{\theta} | R)$. Hence initial samples of the Markov chain should be treated as `burn-in', that is, only the distribution of $\svec{\theta}^{(k)}$ for $k$ exceeding a threshold should be considered. In this article, we discard the first half of the $\svec{\theta}^{(k)}$ for this reason.

To determine the number of MCMC steps needed for the sample distribution of $\svec{\theta}^{(k)}$ to converge to the target posterior $p(\svec{\theta} | R)$, the Gelman--Rubin convergence test (\cite{gelman1992, brooks1998}, also section 11.4 of \cite{gelman2013}), which compares multiple chains of $\svec{\theta}^{(k)}$ under different initial conditions $\svec{\theta}^{(0)}$, is performed. In this article the convergence of the sample distribution to the target is said to have achieved when $\hat{R}$ (as defined in (11.4) of \citet{gelman2013}) corresponding to each component of $\svec{\theta}$ falls below $1.2$.

\subsection{Tuning}
To sample the distribution of $p(\svec{\theta} | R)$ efficiently, the variance of the proposal distribution $V_j$ needs to be tuned. A small variance  $V_j$ leads to successive $\svec{\theta}^{(k)}$ that are very close to one another, while a large $V_j$ leads to numerous rejections; in both cases the support of  $p(\svec{\theta} | R)$ is explored too slowly. For an optimal algorithmic efficiency, a common practice is to maintain the fraction of the candidates $\svec{\vartheta}$ being accepted to be approximately $0.25$ \citep{roberts1997}. Note that this is measured only after the `burn-in' phase. 
A table listing the initial values for the parameter $\svec{\theta}^{(0)}$ and the proposal standard deviation $\sqrt{V_j}$ (before tuning) is given in table \ref{tab:mh_init}.
\begin{table}
  \begin{centering}
  \begin{tabular}{c c c}
     \hline
    Parameter $\svec{\theta}$ & Initial value $\svec{\theta}^{(0)}$ & Proposal standard deviation $\sqrt{V_j}$ \\
    \hline \hline
    $U_0$ & 0~m~s${}^{-1}$ & 0.001~m~s${}^{-1}$ \\
    $\phi_0$ & 0~rad & 0.05~rad \\
    \hline
    $\nu_1$ & 0~s${}^{-1}$ & $2.5 \times 10^{-8}$~s${}^{-1}$ \\
    $\nu_2$ & 0~s${}^{-1}$ & $2.5 \times 10^{-8}$~s${}^{-1}$ \\
    $\phi_\tensor{A}$ & 0~rad & 0.05~rad \\
    \hline
    $\gamma_1$ & 1000~m${}^{2}$~s${}^{-1}$  & 100~m${}^{2}$~s${}^{-1}$ \\
    $\gamma_2$ & 500~m${}^{2}$~s${}^{-1}$  & 50~m${}^{2}$~s${}^{-1}$ \\
    $\phi_\tensor{K}$ & 0~rad & 0.05~rad \\
   \hline
  \end{tabular}
  \caption{
    Parameters used to initialise the  Metropolis--Hastings Algorithm in the `burn-in' phase.}\label{tab:mh_init}
  \end{centering}
\end{table}

The parameter $\svec{\theta}_{\mathrm{MAP}}$ that maximises the posterior density $p(\svec{\theta} | R)$ is used as the initial conditions for tuning $V_j$ and post-`burn-in' sampling. To tune $V_j$, the algorithm is re-run with an additional $8000$ steps, during which the fraction of $\svec{\theta}^{(k)}$ accepted is recorded. If the acceptance fraction exceeds $0.35$, $V_j$ is multiplied by $4/3$; if it is lower than $0.15$, $V_j$ is multiplied by $2/3$. The tuning process is repeated for up to 20 times and stops once the acceptance fraction falls in the range of $[0.15, 0.35]$, in the neighbourhood of the advised value $0.25$ \citep{roberts1997}.
With the tuned variance $V_j$ the  Metropolis--Hasting algorithm is re-run with initial condition $\svec{\theta}_{\mathrm{MAP}}$ and the samples of $\svec{\theta}^{(k)}$ are used for  inference. 

\section{Calculating the Davis diffusivity}\label{app:davis_cal}
The along-stream and cross-stream \citet{davis1987} diffusivities are calculated using the $10$-year Eulerian mean flow at the centre of each cell to define the mean velocity $\bar{\vec{u}} \left( \vec{x} \right)$ appearing in equation \eqref{eqn:kappa_davis}. Evaluating the integral in \eqref{eqn:kappa_davis} requires high temporal resolution; we use  particle locations 
observed every $3$~hours over $10$~years, for $10,000$ particles initially
deployed uniformly across the domain. We adopt the method of \citet{griesel2010} to evaluate the two diffusivities in each of the $16 \times 16$ cells
partitioning the domain. The position of each particle every $3$~hours is
treated as a new independent starting point, to generate a set of particle
trajectories each with time lag $\tau$. The conditional averaging operator $\left< \cdot
\right>_{ \{ \vec{X}_i\left( t \right) = \vec{x} \} }$ in  \eqref{eqn:kappa_davis} is then modified to include
all  particle trajectories that end in a given cell, 
and the time integral is
computed using the trapezoidal rule. Note that, while this formally computes a diffusivity tensor, this tensor need not be symmetric positive
definite (or even symmetric) and hence corresponding diffusivity ellipses cannot
be shown without further processing. Projecting the diffusivity tensor onto directions parallel to and perpendicular to the Eulerian mean flow yields the along-stream and cross-stream diffusivities shown in figure \ref{fig:QGM2_DStemp_KDavis_diffSIv_Idx16_LocLinINC}. 

\bibliographystyle{elsarticle-harv} 
\bibliography{bibliography}

\end{document}